\documentclass[conference]{IEEEtran}
\IEEEoverridecommandlockouts
\usepackage{cite}
\usepackage{amsmath,amssymb,amsfonts}
\usepackage{algorithmic}
\usepackage{graphicx}
\usepackage{textcomp}
\usepackage{xcolor}
\def\BibTeX{{\rm B\kern-.05em{\sc i\kern-.025em b}\kern-.08em
    T\kern-.1667em\lower.7ex\hbox{E}\kern-.125emX}}

\usepackage{graphicx}
\usepackage{caption}
\usepackage{subcaption}
\usepackage{float}
\usepackage{multirow}
\usepackage{amsmath}
\usepackage{nicefrac}  
\usepackage{comment}
\usepackage{xfrac}

\begin{document}

\title{Addressing Limited Data in Auditory Attention Decoding with Diffusion Generative Models*
\thanks{This paper’s data analysis is based on a Master’s Thesis work \cite{rangun2024}}
\thanks{Thanks to the ELLIIT strategic research programme for funding.}

\author{David Rannaleet$^{1,\dagger}$, Victor Gunnarsson$^{1,\dagger}$, Bo Bernhardsson$^{1}$, \\ Martin A. Skoglund$^{2,3,\ddagger}$, and Emina Alickovic$^{2,3,\ddagger}$
\thanks{$^{1}$Department of Automatic Control, Lund University, Lund, Sweden. {\tt\small bob at lth.lu.se }}
\thanks{$^{2}$Eriksholm Research Centre, Snekkersten, Denmark. {\tt\small \{mnsk,eali,dowe\} at eriksholm.com}}%
\thanks{$^{3}$Department of Electrical Engineering, Linköping University, Linköping, Sweden. {\tt\small \{martin.skoglund,emina.alickovic\} at liu.se}}
\thanks{$^{\dagger}$Equally contributed as first authors.}%
\thanks{$^{\ddagger}$Equally contributed as last authors.}%
}%

}

\maketitle

\begin{abstract}
Limited training data constrains deep learning models for Auditory Attention Decoding (AAD) in hearing aids (HAs). AAD uses electroencephalogram (EEG) data to decode listener's attention, enabling real-time tracking of specific sound sources. However, achieving high AAD performance with short time windows typical in HAs ($\leq1$s) is challenging due to the scarcity of real-world speech-evoked EEG data. To address this issue, we investigate diffusion probabilistic models (DPMs) for generating synthetic speech-evoked EEG data. DPMs learn the underlying complex data structure through a denoising process and can generate realistic samples suitable for data augmentation. We evaluate the use of synthetic EEG data for augmenting datasets in locus‑of‑attention (LoA) classification tasks. Our experiments demonstrate that DPMs can generate realistic EEG signals and that incorporating synthetic data significantly improves AAD performance compared to models trained solely on measured EEG data  ($p<.05$). These results highlight the potential of diffusion‑based data augmentation to mitigate training data limitations and improve the robustness of short‑window AAD models in HA applications.
\end{abstract}


\section{Introduction}
Deep learning has advanced hearing aids (HAs), aiming to improve performance in challenging listening environments with competing sound sources \cite{Andersen2021}. However, existing HA amplification strategies often fall short in such scenarios. Auditory attention decoding (AAD)  emerges as a promising personalized solution \cite{o2015attentional, alickovic2019, GeirnaertAttentionDecoding}. AAD uses brain responses to speech, recorded with an electroencephalogram (EEG), to decode the listener's attention, potentially enabling real-time tracking of desired sound sources in HAs. Despite recent advances demonstrating the effectiveness of EEG-based AAD and its potential for non-intrusive, real-time HA enhancement, a key challenge remains: achieving reliable decoding with short time windows (one second or less) typical for HAs. This limitation is largely due to insufficient training data for deep learning models. Speech‑evoked EEG recordings are often limited in duration, as most studies rely on short sessions per subject, and larger, more diverse datasets are needed to improve generalization across listening conditions.

Generative learning provides a partial solution to this data scarcity challenge by learning  patterns in speech-evoked EEG and generating realistic, task-specific signals. Such synthetic data can augment existing datasets, improving AAD performance when real-world data collection is constrained.

Diffusion probabilistic models (DPMs), a class of latent variable generative models, are particularly well suited for this task. By learning to reverse a gradual noise-adding process, DPMs generate realistic samples \cite{sohl2015deep, ho2020inproceeding, song2020score}. Their success in image synthesis and natural language processing suggests strong potential for EEG-based AAD. In this context, DPMs offer two main advantages: they can model complex, non-linear relationships between EEG and auditory stimuli, and they can learn task-specific EEG patterns associated with attention. Although diffusion probabilistic models have previously been applied to EEG-based tasks such as sleep stage classification and cognitive decline detection using long analysis windows (e.g., 30s) \cite{aristimunha2023synthetic, sharma2023medic}, real-time auditory attention decoding requires reliable inference from short time windows ($\leq1$s), motivating generative models capable of capturing informative short-term EEG representations.
 
This work addresses limited EEG data for training AAD models by focusing on attended‑talker classification in multi‑talker environments. We consider locus‑of‑attention (LoA) classification, which decodes attended speech direction (left vs.\ right), and propose a DPM‑based approach to generate synthetic speech‑evoked EEG data. We evaluate DPM‑based data augmentation for LoA tasks and demonstrate improved AAD performance, supporting more accurate and robust real‑time attention decoding in future HAs

\section{Methodology}
\subsection{Experimental Design}
\label{chap:dataset} 

\paragraph{Dataset} This study uses an EEG data set  previously analyzed in \cite{alickovic2021,Andersen2021, tanveer2024, rangun2024} using different methods. 
The dataset comprises recordings from 34 participants (24 male) aged 21-84 (mean: 64.2 years, SD: 13.6 years) with symmetrical sensorineural hearing loss (average 4-frequency pure-tone audiometry of 47.5 dB). All participants were experienced HA users with no reported neurological disorders, dyslexia, or diabetes. The study was approved by the ethics committee of the Capital Region of Denmark (journal no. H-1-2011-033), and written informed consent was obtained. 

\paragraph{Recordings} EEG data were recorded at 1024~Hz using the BioSemi Active Two recording system  with 64 electrodes (international 10-20 system) and two mastoid reference electrodes. 
Participant were seated in a sound-proofed booth facing a screen. The experimental setup consisted of six loudspeakers arranged around the participant, positioned at $\pm 30^\circ$ in front, $\pm 112.5^\circ$ laterally, and $\pm 157.5^\circ$ behind.

\paragraph{Stimuli} The experiment involved a two-talker listening task with background noise. Danish news clips of neutral content spoken by male and female speakers were presented from the front loudspeakers, while background noise simulating a 16-talker environment was presented from rear loudspeakers at -3dB relative to the foreground speech.  

\paragraph{Trials}  Participants completed 80 experimental trials (excluding familiarization trials) across four sessions using a 2×2 hearing‑aid configuration, in which two different noise reduction schemes (NR) were each evaluated with NR enabled or disabled. Each trial comprised a 5‑second background‑noise onset followed by 33 seconds of concurrent target speech and background noise, yielding a total duration of 38 seconds. Within each session, trials were organized into blocks of five with fixed target speaker direction and gender, while speech content was randomized across trials. Participants were instructed to attend to the target speaker and completed a content‑related two‑choice question after each trial. 

\paragraph{Preprocessing} Trials were preprocessed to remove noisy and irrelevant (ocular, muscular, and cardiac) component using band‑pass (0.5–70Hz) and notch (49–51Hz) filtering, downsampling from 1024Hz to 256Hz, visual inspection with channel interpolation, and independent component analysis \cite{keding2024effect}. Data from 31 participants remained after excluding recordings with persistent artifacts. 

\paragraph{Data Split and Normalization} The final dataset comprises 2420 trials of 33\,s, segmented into 1\,s EEG samples using a sliding window, yielding a balanced set of left‑ and right‑target labels (1210 each). Data were split into training (60\%), validation (20\%), and testing (20\%) sets by assigning the first three trials within each block to training, the fourth to validation, and the fifth to testing, ensuring balanced class distributions across splits \cite{tanveer2024}. EEG data were standardized using the mean and standard deviation computed from the training set across all channels.



\subsection{EEG Diffusion Modeling}

DPMs provide an effective approach to data augmentation by gradually adding noise to the data during a forward diffusion process and learning to reverse it through denoising. In forward diffusion, noise is gradually added by progressing through  Markov chain via multiplication with the Markov transition kernel
$
    q(\mathbf{x}_{1:T}|\mathbf{x}_0) = \prod_{t=1}^{T}q(\mathbf{x}_{t}|\mathbf{x}_{t-1})
$
with $t=0$ denoting the original data distribution $q(\mathbf{x}_0)$ and $t=T$ is the final instance of the chain representing the converted data distribution $q(\mathbf{x}_t)$. The reverse process, as defined by \cite{sohl2015deep}, is 
$
    p_\theta(\mathbf{x}_{0:T}) = p(\mathbf{x}_T)\prod_{t=1}^{T}p_\theta(\mathbf{x}_{t-1}|\mathbf{x}_{t})$,
where $p_\theta(\mathbf{x}_{t-1}|\mathbf{x}_{t})$ is the reverse Markov transition kernel.
The Gaussian forward kernel is defined as
  $   q(\mathbf{x}_t|\mathbf{x}_{t-1}) =  \mathcal{N}(\mathbf{x}_t; \sqrt{1- \beta_t}\mathbf{x}_{t-1}, \beta_t \mathit{I}),$ where $\beta_t$ is the variance of the introduced noise. Given a small size of the variance $\beta_t$, the reverse Markov transition kernel will be of the same functional form as the forward process: $
    p_\theta(\mathbf{x}_{t-1}|\mathbf{x}_{t}) = \mathcal{N}(\mathbf{x}_{t-1}; \mu_{\theta}(\mathbf{x}_t,t), \Sigma_{\theta}(\mathbf{x}_t,t)).$
Here, the mean $\mu_{\theta}(\mathbf{x}_t,t)$ and covariance $\Sigma_{\theta}(\mathbf{x}_t,t)$ are unknown parameters, which are estimated by a neural network.

\paragraph{Implicit Diffusion}
We specifically selected Denoising Diffusion Implicit Models (DDIMs) \cite{song2022denoising} due to their ability 
to generate new samples without requiring sequential time steps, unlike e.g. Denoising Diffusion Probabilistic Model (DDPM). 
In DDIM, we use the forward distributions indexed by $\sigma$,
 $   q_\sigma(\mathbf{x}_{1:T}|\mathbf{x}_t, \mathbf{x}_0) = q_\sigma(\mathbf{x}_T|\mathbf{x}_0) \prod_{t=2}^T q_\sigma(\mathbf{x}_{t-1}|\mathbf{x}_t, \mathbf{x}_0 )$
where $q_\sigma(\mathbf{x}_{t-1}|\mathbf{x}_t, \mathbf{x}_0) = \mathcal{N} \left ( \sqrt{\Bar{\alpha}_{t-1}}\mathbf{x}_0 + \sqrt{1-\Bar{\alpha}_{t-1} - \sigma_t^2} \cdot \sfrac{\mathbf{x}_t-\sqrt{\Bar{\alpha}_t \mathbf{x}_0}}{\sqrt{1-\Bar{\alpha}_{t-1}}}, \sigma_t^2 \mathit{I} \right)$, with $\alpha$ and its cumulative product as $\Bar{\alpha}$, $\alpha_t = 1 - \beta_t$ and $\Bar{\alpha_t} = \prod_{s=1}^{t} \alpha_s$. 
The DDIM training loop is identical to that of DDPM, as it models the same parameter $\epsilon_\theta$ and uses the same forward diffusion for adding noise.
Since the reverse diffusion process is determined by the forward diffusion process, it is possible to use another forward diffusion $q_\sigma$ defined on a subset of latent variables $\mathbf{x}_{\tau_1}, \dots, \mathbf{x}_{\tau_S}$ during sampling, thus reducing the number of steps from $T$ to $S$ as
  $  \mathbf{x}_{\tau_{i-1}}(\eta) = \sqrt{\Bar{\alpha}_{\tau_{i-1}}} \left( \sfrac{\mathbf{x}_{\tau_i} - \sqrt{1-\Bar{\alpha}_{\tau_i}}\epsilon_\theta^{(\tau_i)}(\mathbf{x}_{\tau_i})}{\sqrt{\Bar{\alpha}_{\tau_i}}} \right)  + \sqrt{1-\Bar{\alpha}_{\tau_{i-1}}- \sigma_{\tau_i}(\eta)^2} \epsilon_\theta^{(\tau_i)}(\mathbf{x}_{\tau_i}) + \sigma_{\tau_i}(\eta) \epsilon_i$
where $\sigma_{\tau_i}(\eta) = \eta\sqrt{ \sfrac{1- \Bar{\alpha}_{\tau_{i-1}}}{1- \Bar{\alpha}_{\tau_{i}}}} \sqrt{ 1-\sfrac{\Bar{\alpha}_{\tau_{i}}}{\Bar{\alpha}_{\tau_{i-1}}}}$ and $\eta$ is a hyperparameter that sets the stochasticity of the process. 
The variances ($\beta$) of the diffusion model are set when initializing the DDIM schedule using a $\beta$ scheduler. Our implementation uses the squared cosine scheduler proposed by \cite{alex2021}. Each $\beta_t$ is now set to the value of
 $   \beta_t = 1 - \sfrac{\Bar{\alpha_t}}{\Bar{\alpha}_{t-1}}$
, where 
 $   \Bar{\alpha_t} = \sfrac{f(t)}{f(0)}, f(t)= \cos{\left( \sfrac{t/T + s}{1+s} \cdot \sfrac{\pi}{2} \right)^2.}$
The initial value $\mathbf{x}_0 \sim \left( \sfrac{\mathbf{x}_{\tau_i} - \sqrt{1-\Bar{\alpha}_{\tau_i}}\epsilon_\theta^{(\tau_i)}(\mathbf{x}_{\tau_i})}{\sqrt{\Bar{\alpha}_{\tau_i}}} \right)$ is updated for each step in the sampling process. 

\paragraph{U-Net}
We use a variant of the original U‑Net architecture \cite{ronneberger2015unet}, commonly used for image domain diffusion. The U-Net takes a noisy EEG sample and corresponding diffusion time step as inputs, and outputs a  of the mean $\mu_\theta(\mathbf{x}_t,t)$ in the reverse diffusion kernel $p_\theta(\mathbf{x}_{t-1}|\mathbf{x}_{t})$. The implementation is adapted from the HuggingFace \textit{diffusers} library, built on \textit{transformers} library \cite{wolf2020huggingfaces}. 
Our U-Net uses a base of six convolution blocks each for the encoder and decoder. All convolutions in the network use a $3 \times 3$ kernel and the Sigmoid-Linear Unit (SiLU) activation function. To provide time‑step information absent in the original U‑Net, we incorporate sinusoidal positional embeddings, following the DDPM formulation \cite{ho2020inproceeding}.

\paragraph{V-prediction}
Diffusion models are commonly trained to predict the added noise $\boldsymbol{\epsilon}$, referred to here as the Epsilon loss.  Alternatively, the model can  predict the sample ($\mathbf{x}$) itself or a combination of both, known as v-prediction (V-pred) \cite{salimans2022vpred}. In V-pred, a new prediction target $\mathbf{v}$ is defined as
 $   \mathbf{v} \equiv \alpha_{t}\epsilon - \sigma_{t} \mathbf{x}$
with the corresponding loss as $||\mathbf{v}-\mathbf{\hat{v}}||^2$.

\paragraph{Spectral Loss}
To consider spectral components of EEG, a spectral loss using the Short-Time Fourier Transform (STFT) was used, which takes the form of: 
 $  \sfrac{1}{N}\| |STFT(\mathbf{x})| - |STFT(\hat{\mathbf{x}}) | \|_2^2$.
Here, $\mathbf{x}$ represents the original data, $\hat{\mathbf{x}}$ is the predicted data and $N$ is the number of elements in $\mathbf{x}$. This approach is inspired by the spectral loss used as a reconstruction loss of a VQ-VAE in \cite{dhariwal2020jukebox} to consider mid- and high frequencies. The data is scaled such that the loss falls within the interval $[0, 0.5]$ to balance it with the existing average MSE loss. 

\paragraph{Jensen-Shannon Distance}
To quantitatively assess the similarity between  real EEG and generated EEG data distributions, the Jensen-Shannon Distance (JSD) was used. It is defined as the square-root of the Jensen-Shannon divergence $\sfrac{1}{2} \cdot D_{KL}(P \| \sfrac{P+Q}{2}) + 
 \sfrac{1}{2} \cdot D_{KL}(Q \| \sfrac{Q+P}{2}), $ where $P$ and $Q$ represent two distributions being compared, and $D_{KL}(\cdot)$ is the Kullback-Liebler (KL) divergence. 

\paragraph{Implementation}
To reduce sensitivity to STFT parameter choices, the mean is taken across multiple STFT resolutions, preventing overfitting to one STFT representation and allowing the model to capture a broader range of time-frequency structure \cite{yamamoto2020parallel}.
Loss implementations follow a modified version of the code from \cite{steinmetz2020auraloss}.
Training uses $T = 1000$ diffusion steps  \cite{sohl2015deep,ho2020inproceeding}. Input samples are one-second EEG segments with a shape of [$N$,1,64,256] and a batch size of $N=64$. The model is trained for 100 epochs using the AdamW optimizer \cite{hutter2019decoupled}, an initial learning rate of 1e-4 and a cosine schedule with 500 warmup steps. In our testing, 100 epochs were sufficient for loss convergence.

To prevent sample saturation, dynamic thresholding is applied \cite{saharia2022photorealistic}. At each diffusion step a percentile‑based threshold $s$ is computed and values beyond $\pm s$ are clipped. Because the EEG data are standardized rather than normalized, we cap the threshold at \(s_{\max}=5\) (i.e., if \(s > s_{\max}\), we use \(s_{\max}\)) and do not rescale after clipping, thereby bounding values to \([-5,5]\) and suppressing extreme outliers. Separate diffusion models are trained for left- and right-attention labels, and 45{,}000 one-second EEG samples are generated per label. Full training details are provided in our GitHub repository.\footnote{Link provided upon manuscript decision.}

\subsection{Locus of Attention Classification}
\label{sec:loa}

AAD aims to identify the attended speaker in multi‑speaker environments, with locus‑of‑attention (LoA) methods forming a subset that decode the attended speech direction (left vs.\ right) from EEG  \cite{servaas2021, wilroth2023improving, Puffay_2023, tanveer2024}.
For LoA classification, we employ the EEGNeX classifier \cite{CHEN2024105475}, selected for its strong performance across EEG classification tasks compared to earlier convolutional neural network models such as EEGNet \cite{Lawhern2018}. We use a PyTorch reimplementation of the original EEGNeX architecture \cite{anselpyTorch2024}.

Classifiers were trained with a batch size of 64 for 100 epochs using the AdamW optimizer \cite{hutter2019decoupled} and a fixed learning rate of $5\times10^{-4}$. EEG data were sampled using 1‑second sliding windows with a stride of 0.75\,s, resulting in a 25\% overlap between consecutive samples and no overlap across trials. Data were split into training (60\%), testing (20\%), and validation (20\%) sets, and for each configuration, the model with the lowest test‑set loss was evaluated on the held‑out validation set.

To assess the impact of synthetic EEG data for data augmentation, we first trained a baseline classifier and reused its hyperparameters across all models for consistency. Augmented models combined real EEG training data with synthetic EEG at different ratios (15\%, 30\%, 60\%, or 100\% of the original training set size), and each configuration was trained 20 times to estimate average performance and confidence intervals. Since diffusion models generate data from normally distributed noise, we included a noise addition model as a simpler augmentation baseline, in which Gaussian noise sampled from 
\(\mathcal{N}(0, 0.15)\) was added to each data point of each EEG channel. The standard deviation of the noise was empirically tuned by visually inspecting samples before and after noise addition so that the original signal remained identifiable. This noise addition was applied to 15\% of the training data.



\section{Results and Discussion}
\subsection{Diffusion Model Performance}

Directly computing JSD on high‑dimensional EEG data is computationally demanding; therefore, EEG distributions were approximated using channel‑wise histograms.  For each attention label (Left or Right), 15{,}000 random samples were drawn from standardized real and generated EEG data. Per‑channel histograms were computed using 200 equal‑width bins over the range $\left[-10,10\right]$, and the resulting distributions ($P$ and $Q$) were used to compute JSD.

Across diffusion model losses (Epsilon, V‑pred, and Spectral) and attention targets, JSD values were comparable. Epsilon showed JSD values of 0.026 (Left) and 0.028 (Right); V‑pred showed higher JSD values (0.064 and 0.072); and Spectral  had intermediate JSD values of 0.042 (Left) and 0.047 (Right). These results indicate limited dependence of distributional similarity on the specific diffusion configuration, with Epsilon and Spectral models showing closely matched similarity to real EEG data.

\begin{figure}[!htbp]
    \centering
    \includegraphics[width=\linewidth]{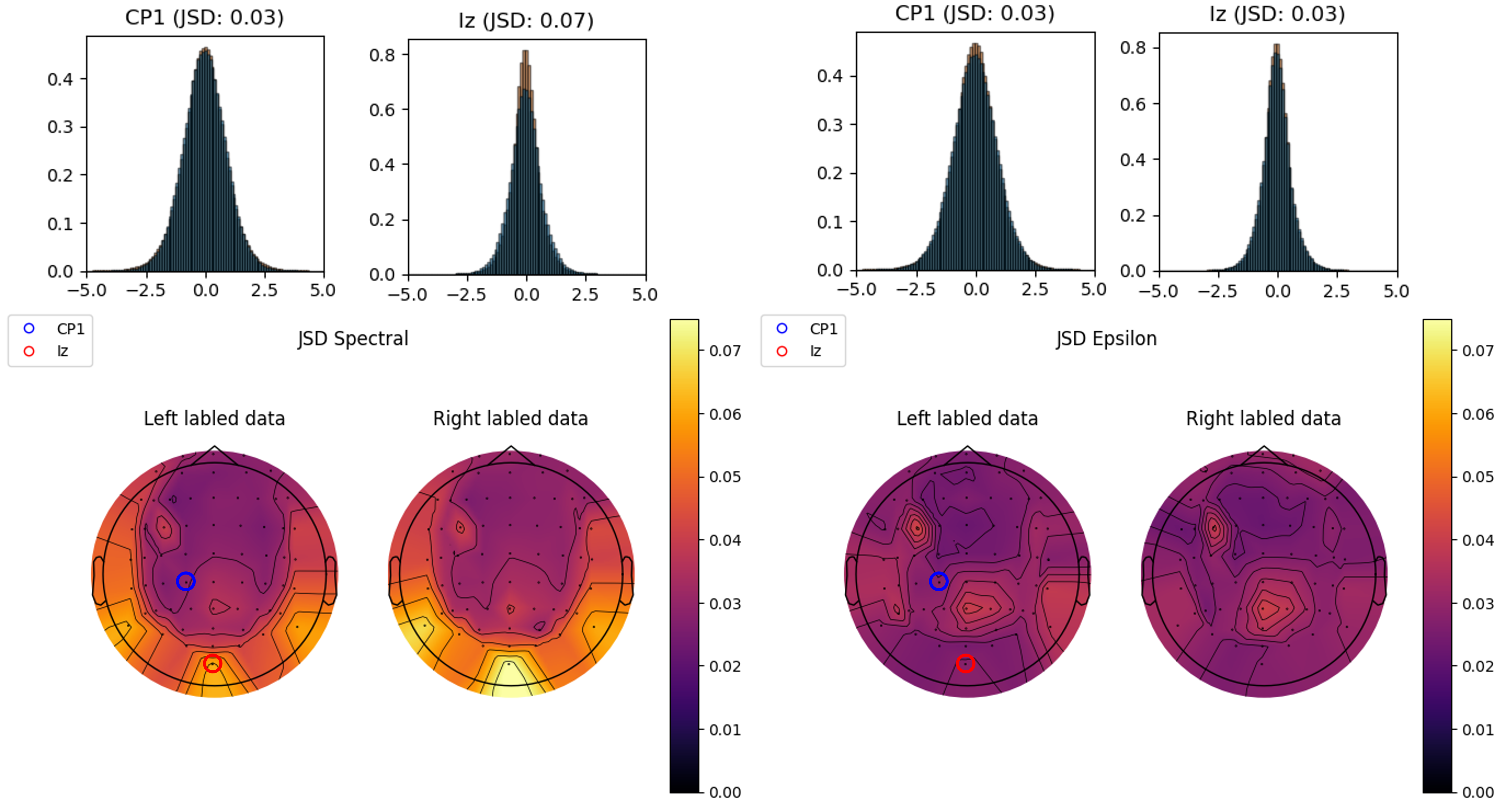}
    \caption{JSD between real and synthetic EEG for two models. The topoplot shows channel‑wise JSD, and two channels are highlighted with corresponding histograms (red: real, blue: synthetic). The largest differences appear near zero.  
    }
    \label{fig:JSDtopoplot}
\end{figure}

Figure~\ref{fig:JSDtopoplot} shows  channel‑wise JSD for the Epsilon and Spectral models. Two specific channels are highlighted with corresponding histograms.  Most channels exhibit low divergence, with fronto‑central regions showing particularly high similarity between real and synthetic EEG. Most channels show minimal JSD variations, making the differences visually subtle; slight deviations occur near zero, where real EEG displays higher density. This spatial pattern corresponds to known attention‑related cortical activity \cite{giard1988several, crosse2015congruent, etard2019neural}, suggesting that the generative models capture task‑relevant EEG structure. 


\subsection{Classification}

\begin{figure}[!htb]
    \centering
    \includegraphics[width=\linewidth]{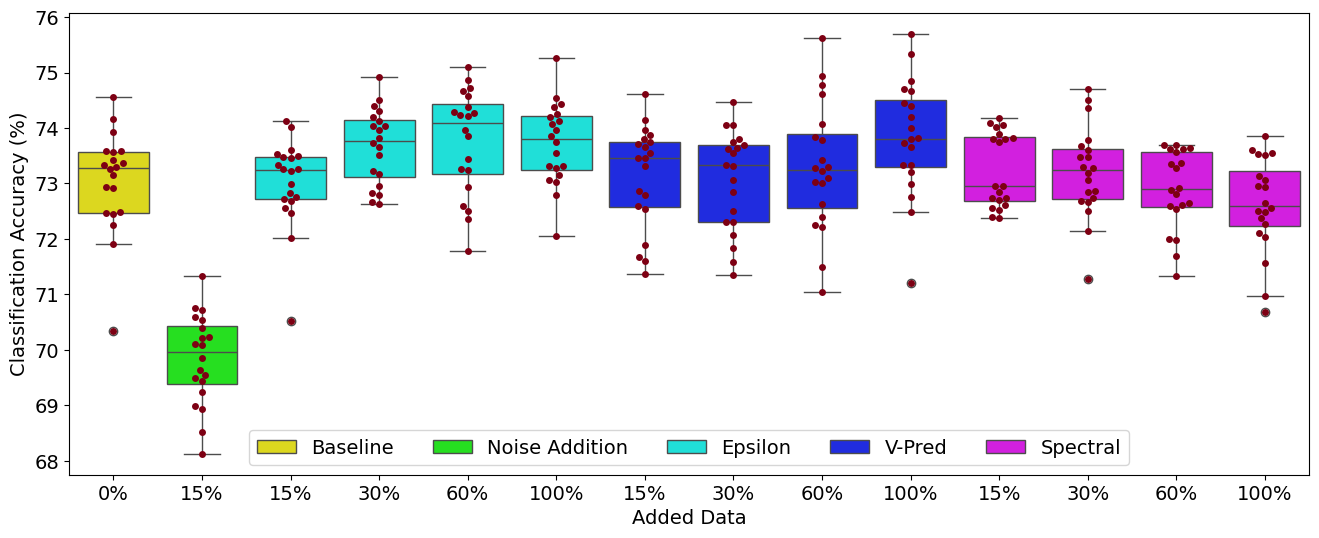}
    \caption{Box‑plot of LoA classification accuracy across models. Each dot shows a single run; circled points indicate outliers beyond $\pm 1.5$ times the interquartile range.}
    \label{fig:boxplot_classifier_LvR_VALIDATION}
\end{figure}

Figure~\ref{fig:boxplot_classifier_LvR_VALIDATION} compares LoA classification accuracy between the baseline, the 'noise addition' model, and the  diffusion models. The boxplot shows the interquartile range containing 50 \% of the data, with a median marked by a line. Data outside the whiskers are considered outliers. The 'noise addition' model shows a drop of around 3\% relative to the baseline, indicating that adding noise alone does not improve performance. Most diffusion models perform similarly to the baseline, with some achieving modest but statistically supported improvements.

To assess statistical differences relative to the baseline, we applied Dunnett’s test \cite{dunnett1955multiple}. Table~\ref{tab:t_test_LvR_VALIDATION} summarizes the resulting $p$‑values. Three configurations differed from the baseline: the noise‑addition model showed significantly lower accuracy ($p<10^{-4}$), whereas the Epsilon model with 60\% synthetic data ($p=0.083$) and the Spectral model with 100\% synthetic data ($p=0.042$) achieved higher accuracy.

\begin{table}[h]
\centering
\caption{P-values from Dunnett's test vs. baseline ($n=20$). $\ll\mathbf{0.01}$ denotes $p<10^{-4}$.}
\begin{tabular}{lcccc}
\hline
\textbf{Model} & \textbf{15\%} & \textbf{30\%} & \textbf{60\%} & \textbf{100\%} \\ \hline\hline
Epsilon (LoA)   & 1.0 & 0.18 & \textbf{0.083} & 0.12 \\
Spectral (LoA)  & 1.0 & 1.0  & 0.98 & \textbf{0.042} \\
V-pred (LoA)    & 1.0 & 1.0  & 1.0  & 0.61 \\
Noise Addition  & \multicolumn{4}{c}{$\ll \mathbf{0.01}$ (15\% only)} \\ \hline
\end{tabular}
\label{tab:t_test_LvR_VALIDATION}
\end{table}

Table~\ref{tab:t_test_LvR_improvement_VALIDATION} reports the accuracy improvements for the two diffusion‑based configurations with $p<0.10$. The Epsilon model with 60\% synthetic data improves accuracy by 0.71\%, and the Spectral model with 100\% synthetic data improves accuracy by 0.78\%. Although the confidence intervals overlap, indicating no clear difference between the two configurations, both outperform the baseline.

\begin{table}[!htpb]
\centering
\caption{Mean improvement (MI, $n=20$) in accuracy (Acc) for the models with significant P-values ($< 0.10$). The confidence interval (CI) refers to the range of MI.}
\begin{tabular}{ccccc}
\hline
\textbf{Model} & \textbf{Added data (\%)} & \textbf{Acc (\%)} & \textbf{MI (\%)} & \textbf{95\% CI} \\ \hline\hline
Baseline & 0 & 73.05 & N/A & N/A \\ \hline
Epsilon & 60 & 73.76 & 0.71 & 0.12 to 1.31 \\ \hline
Spectral & 100 & \textbf{73.83} & \textbf{0.78} & 0.16 to 1.41 \\ \hline
\end{tabular}
\label{tab:t_test_LvR_improvement_VALIDATION}
\end{table}

\label{sec:discussion}
This study relies on a single EEG dataset, which limits generalizability. However, our methods do not depend on dataset‑specific features, suggesting potential applicability to other datasets. Future work should evaluate additional EEG corpora and recording conditions. The analysis also assumes that trial‑level labels accurately reflect attention and that each 1\,s segment contains sufficient information about the attended talker. Occasional label mismatches are possible and may affect training and evaluation.

Computational constraints limited the number of classifier configurations explored, and only 20 runs per setting were performed for statistical comparisons. Although diffusion models are computationally intensive to train, once trained they generate synthetic EEG much faster than acquiring equivalent real data, partially mitigating this limitation.

We did not assess whether diffusion models may reproduce near‑duplicate EEG samples, a phenomenon observed in other diffusion domains \cite{Somepalli2022}. Duplication could introduce bias or reduce generalization and should be examined in future work. Finally, as synthetic EEG becomes more realistic, safeguards may be needed to prevent generated signals from being misinterpreted as clinical data, and clear usage guidelines may help mitigate this risk.

\section{Conclusion}
\label{sec:conclusion}

This work assessed the use of DPMs for generating synthetic speech‑evoked EEG to support AAD. The generated signals improved classification of attended‑speech direction compared to models trained on measured EEG alone ($p<.1$), indicating that DPM‑based augmentation can help address data limitations in AAD.

Several directions remain for future work. Further tuning of DPM parameters is needed, as this was not addressed in detail. Replacing the U-Net neural network architecture with a model tailored for multichannel time-series data could be a promising direction. While U-Net is effective for image-based diffusion models, it is likely not optimally suited for EEG data. Developing an inner denoising model that better differentiates noise from distinctive EEG features could improve generation quality. Such a model would need to effectively capture the temporal and spatial dynamics inherent in EEG signals.

Furthermore, exploring conditional and multimodal diffusion models might improve the realism and task-specificity of the generated EEG data. Developing subject-specific models that can adapt to individual EEG patterns is another possible direction for more personalized AAD applications. Finally, adapting the diffusion model to handle continuous EEG data of arbitrary lengths, both as input and output, could provide much greater task flexibility. This approach would enable the generation of longer, more coherent sequences that better mimic real EEG signals.

\section*{DECLARATION OF GENERATIVE AI AND AI-ASSISTED TECHNOLOGIES}
During the preparation of this work, the authors used ChatGPT‑5.1 to improve readability through rephrasing and grammar checking. The authors reviewed and edited the content as required and take full responsibility for the publication.

\bibliographystyle{IEEEtran}

\raggedbottom
\bibliography{referencesEUSIPCO}

@misc{rangun2024,
  author       = {{Rannaleet, David and Gunnarsson, Victor}},
  language     = {{eng}},
    title        = {{Diffusion Modelling approaches to EEG-based Auditory Attention Decoding}},
  year         = {{2024}},
school={{M}aster’s {T}hesis, Lund University},
note={{M}aster’s {T}hesis, Lund University}
}

@inproceedings{keding2024effect,
  title={Effect of Independent Component Artifact Rejection on EEG-Based Auditory Attention Decoding},
  author={Keding, Oskar and Wilroth, Johanna and Skoglund, Martin A and Alickovic, Emina},
  booktitle={2024 32nd European Signal Processing Conference (EUSIPCO)},
  pages={877--881},
  year={2024},
  organization={IEEE}
}

@inproceedings{aristimunha2023synthetic,
  title={{Synthetic Sleep EEG Signal Generation using Latent Diffusion Models}},
  author={Aristimunha, Bruno and de Camargo, Raphael Yokoingawa and Chevallier, Sylvain and Lucena, Oeslle and Thomas, Adam G and Cardoso, M Jorge and Pinaya, Walter Hugo Lopez and Dafflon, Jessica},
  booktitle={Deep Generative Models for Health Workshop NeurIPS 2023},
  year={2023}
}

@inproceedings{sharma2023medic,
  title={{MEDiC: Mitigating EEG Data Scarcity Via Class-Conditioned Diffusion Model}},
  author={Sharma, Gulshan and Dhall, Abhinav and Subramanian, Ramanathan},
  booktitle={Deep Generative Models for Health Workshop NeurIPS 2023},
  year={2023}
}

@article{GramfortEtAl2013a,
  title = {{{MEG}} and {{EEG}} Data Analysis with {{MNE}}-{{Python}}},
  author = {Gramfort, Alexandre and Luessi, Martin and Larson, Eric and Engemann, Denis A. and Strohmeier, Daniel and Brodbeck, Christian and Goj, Roman and Jas, Mainak and Brooks, Teon and Parkkonen, Lauri and H{\"a}m{\"a}l{\"a}inen, Matti S.},
  year = {2013},
  volume = {7},
  pages = {1--13},
  doi = {10.3389/fnins.2013.00267},
  journal = {Frontiers in Neuroscience},
  number = {267}
}

@article{giard1988several,
  title={Several attention-related wave forms in auditory areas: a topographic study},
  author={Giard, MH and Perrin, F and Pernier, J and Peronnet, F},
  journal={Electroencephalography and Clinical Neurophysiology},
  volume={69},
  number={4},
  pages={371--384},
  year={1988},
  publisher={Elsevier}
}

@article{crosse2015congruent,
  title={Congruent visual speech enhances cortical entrainment to continuous auditory speech in noise-free conditions},
  author={Crosse, Michael J and Butler, John S and Lalor, Edmund C},
  journal={Journal of Neuroscience},
  volume={35},
  number={42},
  pages={14195--14204},
  year={2015},
  publisher={Soc Neuroscience}
}

@article{etard2019neural,
  title={Neural speech tracking in the theta and in the delta frequency band differentially encode clarity and comprehension of speech in noise},
  author={Etard, Octave and Reichenbach, Tobias},
  journal={Journal of Neuroscience},
  volume={39},
  number={29},
  pages={5750--5759},
  year={2019},
  publisher={Soc Neuroscience}
}

@article{GeirnaertAttentionDecoding,
Author = {Geirnaert, Simon and Vandecappelle, Servaas and Alickovic, Emina and de Cheveigne, Alain and Lalor, Edmund and Meyer, Bernd T. and Miran, Sina and Francart, Tom and Bertrand, Alexander},
ISSN = {10535888},
Journal = {IEEE Signal Processing Magazine},
Number = {4},
Pages = {89 - 102},
Title = {Electroencephalography-Based Auditory Attention Decoding: Toward Neurosteered Hearing Devices.},
Volume = {38},
Year = {2021},
}

@inproceedings{hutter2019decoupled,
  title={Decoupled weight decay regularization},
  author={Hutter, Frank and Loshchilov, Ilya},
  booktitle={International Conference on Learning Representations (ICLR)},
  volume={7},
  year={2019}
}

@article {servaas2021,
article_type = {journal},
title = {{EEG}-based detection of the locus of auditory attention with convolutional neural networks},
author = {Vandecappelle, Servaas and Deckers, Lucas and Das, Neetha and Ansari, Amir Hossein and Bertrand, Alexander and Francart, Tom},
editor = {Shinn-Cunningham, Barbara G and O'Sullivan, James and Dimitrijevic, Andrew},
volume = 10,
year = 2021,
month = {4},
pub_date = {2021-04-30},
pages = {e56481},
citation = {eLife 2021;10:e56481},
journal = {eLife},
issn = {2050-084X},
publisher = {eLife Sciences Publications, Ltd},
}

@ARTICLE{alickovic2021,
AUTHOR={Alickovic, Emina and Ng, Elaine Hoi Ning and Fiedler, Lorenz and Santurette, Sébastien and Innes-Brown, Hamish and Graversen, Carina},
TITLE={Effects of Hearing Aid Noise Reduction on Early and Late Cortical Representations of Competing Talkers in Noise},
JOURNAL={Frontiers in Neuroscience},
VOLUME={15},
YEAR={2021},
ISSN={1662-453X}
}

@article{alickovic2019,
Author = {Alickovic, Emina and Lunner, Thomas and Gustafsson, Fredrik and Ljung, Lennart},
ISSN = {1662-453X},
Journal = {Frontiers in Neuroscience},
Title = {A Tutorial on Auditory Attention Identification Methods.},
Volume = {13},
Year = {2019},
}

@article{tanveer2024,
	author={Tanveer, M Asjid and Skoglund, Martin A. and Bernhardsson, Bo and Alickovic, Emina},
	title={Deep learning-based auditory attention decoding in listeners with hearing impairment},
	journal={Journal of Neural Engineering},
	year={2024},
}

@article{Andersen2021,
Author = {Andersen, Asger Heidemann and Santurette, Sébastien and Pedersen, Michael Syskind and Alickovic, Emina and Fiedler, Lorenz and Jensen, Jesper and Behrens, Thomas},
ISSN = {07340451},
Journal = {Seminars in Hearing},
Number = {3},
Pages = {260 - 281},
Title = {Creating Clarity in Noisy Environments by Using Deep Learning in Hearing Aids.},
Volume = {42},
Year = {2021},
}

@article{o2015attentional,
  title={{Attentional selection in a cocktail party environment can be decoded from single-trial EEG}},
  author={O'sullivan, James A and Power, Alan J and Mesgarani, Nima and Rajaram, Siddharth and Foxe, John J and Shinn-Cunningham, Barbara G and et.al.},
  journal={Cerebral cortex},
  volume={25},
  number={7},
  pages={1697--1706},
  year={2015},
  publisher={Oxford University Press}
}

@article{Puffay_2023,
year = {2023},
month = {aug},
publisher = {IOP Publishing},
volume = {20},
number = {4},
pages = {041003},
author = {Corentin Puffay and Bernd Accou and Lies Bollens and Mohammad Jalilpour Monesi and Jonas Vanthornhout and Hugo Van hamme and Tom Francart},
title = {Relating {EEG} to continuous speech using deep neural networks: a review},
journal = {Journal of Neural Engineering}
}

@article{wilroth2023improving,
  title={Improving {EEG}-based decoding of the locus of auditory attention through domain adaptation},
  author={Wilroth, Johanna and Bernhardsson, Bo and Heskebeck, Frida and Skoglund, Martin A. and Bergeling, Carolina and Alickovic, Emina},
  journal={Journal of Neural Engineering},
  volume={20},
  number={6},
  pages={066022},
  year={2023},
  publisher={IOP Publishing}
}

@inproceedings{alex2021,
  title={Improved denoising diffusion probabilistic models},
  author={Nichol, Alexander Quinn and Dhariwal, Prafulla},
  booktitle={International conference on machine learning},
  pages={8162--8171},
  year={2021},
  organization={PMLR}
}

@inproceedings{song2022denoising,
  author       = {Jiaming Song and
                  Chenlin Meng and
                  Stefano Ermon},
  title        = {Denoising Diffusion Implicit Models},
  booktitle    = {9th International Conference on Learning Representations, {ICLR} 2021,
                  Virtual Event, Austria, May 3-7},
  year         = {2021},
  bibsource    = {dblp computer science bibliography, https://dblp.org}
}

@inproceedings{yamamoto2020parallel,
  author       = {Ryuichi Yamamoto and
                  Eunwoo Song and
                  Jae{-}Min Kim},
  title        = {{Parallel WaveGAN: {A} Fast Waveform Generation Model Based on Generative
                  Adversarial Networks with Multi-Resolution Spectrogram}},
  booktitle    = {2020 {IEEE} International Conference on Acoustics, Speech and Signal
                  Processing, {ICASSP} 2020, Barcelona, Spain, May 4-8, 2020},
  pages        = {6199--6203},
  publisher    = {{IEEE}},
  year         = {2020},
  bibsource    = {dblp computer science bibliography, https://dblp.org}
}

@InProceedings{ronneberger2015unet,
author="Ronneberger, Olaf
and Fischer, Philipp
and Brox, Thomas",
editor="Navab, Nassir
and Hornegger, Joachim
and Wells, William M.
and Frangi, Alejandro F.",
title="U-Net: Convolutional Networks for Biomedical Image Segmentation",
booktitle="Medical Image Computing and Computer-Assisted Intervention -- MICCAI 2015",
year="2015",
publisher="Springer International Publishing",
address="Cham",
pages="234--241",
isbn="978-3-319-24574-4"
}

@article{dhariwal2020jukebox,
  title={Jukebox: A generative model for music},
  author={Dhariwal, Prafulla and Jun, Heewoo and Payne, Christine and Kim, Jong Wook and Radford, Alec and Sutskever, Ilya},
  journal={arXiv preprint arXiv:2005.00341},
  year={2020}
}

@inproceedings{steinmetz2020auraloss,
    title={auraloss: {A}udio focused loss functions in {PyTorch}},
    author={Steinmetz, Christian J. and Reiss, Joshua D.},
    booktitle={Digital Music Research Network One-day Workshop (DMRN+15)},
    year={2020}
}

@article{saharia2022photorealistic,
  title={Photorealistic text-to-image diffusion models with deep language understanding},
  author={Saharia, Chitwan and Chan, William and Saxena, Saurabh and Li, Lala and Whang, Jay and Denton, Emily L and Ghasemipour, Kamyar and Gontijo Lopes, Raphael and Karagol Ayan, Burcu and Salimans, Tim and others},
  journal={Advances in neural information processing systems},
  volume={35},
  pages={36479--36494},
  year={2022}
}

@article{CHEN2024105475,
title = {Toward reliable signals decoding for electroencephalogram: A benchmark study to {EEGNeX}},
journal = {Biomedical Signal Processing and Control},
volume = {87},
pages = {105475},
year = {2024},
issn = {1746-8094},
author = {Xia Chen and Xiangbin Teng and Han Chen and Yafeng Pan and Philipp Geyer}
}

@article{Lawhern2018,
   title={{EEGNet}: a compact convolutional neural network for {EEG}-based brain–computer interfaces},
   volume={15},
   ISSN={1741-2552},
   number={5},
   journal={Journal of Neural Engineering},
   publisher={IOP Publishing},
   author={Lawhern, Vernon J and Solon, Amelia J and Waytowich, Nicholas R and Gordon, Stephen M and Hung, Chou P and Lance, Brent J},
   year={2018},
   month=jul, pages={056013} }

@INPROCEEDINGS{Somepalli2022,
  author={Somepalli, Gowthami and Singla, Vasu and Goldblum, Micah and Geiping, Jonas and Goldstein, Tom},
  booktitle={2023 IEEE/CVF Conference on Computer Vision and Pattern Recognition (CVPR)}, 
  title={Diffusion Art or Digital Forgery? Investigating Data Replication in Diffusion Models}, 
  year={2023},
  volume={},
  number={},
  pages={6048-6058},
  doi={10.1109/CVPR52729.2023.00586},
  ISSN={2575-7075},
  month={June},}

@inproceedings{
salimans2022vpred,
title={Progressive Distillation for Fast Sampling of Diffusion Models},
author={Tim Salimans and Jonathan Ho},
booktitle={International Conference on Learning Representations},
year={2022},
}

@inproceedings{anselpyTorch2024,
author = {Ansel, Jason and Yang, Edward and He, Horace and Gimelshein, Natalia and Jain, Animesh and Voznesensky, Michael and Bao, Bin and et.al.},
booktitle = {29th ACM International Conference on Architectural Support for Programming Languages and Operating Systems, Volume 2 (ASPLOS '24)},
doi = {10.1145/3620665.3640366},
month = apr,
publisher = {ACM},
title = {{PyTorch 2: Faster Machine Learning Through Dynamic Python Bytecode Transformation and Graph Compilation}},
year = {2024}
}

@article{dunnett1955multiple,
  title={A multiple comparison procedure for comparing several treatments with a control},
  author={Dunnett, Charles W},
  journal={Journal of the American Statistical Association},
  volume={50},
  number={272},
  pages={1096--1121},
  year={1955},
  publisher={Taylor \& Francis}
}

@inproceedings{wolf2020huggingfaces,
    title = "Transformers: State-of-the-Art Natural Language Processing",
    author = "Wolf, Thomas  and
      Debut, Lysandre  and
      Sanh, Victor  and
      Chaumond, Julien  and
      Delangue, Clement  and
      Moi, Anthony and et.al.",
    editor = "Liu, Qun  and
      Schlangen, David",
    booktitle = "Proceedings of the 2020 Conference on Empirical Methods in Natural Language Processing: System Demonstrations",
    month = oct,
    year = "2020",
    publisher = "Association for Computational Linguistics",
    pages = "38--45",
}

@inproceedings{sohl2015deep,
  title={Deep unsupervised learning using nonequilibrium thermodynamics},
  author={Sohl-Dickstein, Jascha and Weiss, Eric and Maheswaranathan, Niru and Ganguli, Surya},
  booktitle={International conference on machine learning},
  pages={2256--2265},
  year={2015},
  organization={PMLR}
}

@article{song2020score,
  title={Score-based generative modeling through stochastic differential equations},
  author={Song, Yang and Sohl-Dickstein, Jascha and Kingma, Diederik P and Kumar, Abhishek and Ermon, Stefano and Poole, Ben},
  journal={arXiv preprint arXiv:2011.13456},
  year={2020}
}

@inproceedings{ho2020inproceeding,
 author = {Ho, Jonathan and Jain, Ajay and Abbeel, Pieter},
 booktitle = {Advances in Neural Information Processing Systems},
 editor = {H. Larochelle and M. Ranzato and R. Hadsell and M.F. Balcan and H. Lin},
 pages = {6840--6851},
 publisher = {Curran Associates, Inc.},
 title = {Denoising Diffusion Probabilistic Models},
 volume = {33},
 year = {2020}
}

\end{document}